\documentclass[sigconf, screen=true, nonacm=true, anonymous=false]{acmart}
\usepackage{enumerate}
\usepackage{subcaption}
\usepackage{hyperref}
\def\BibTeX{{\rm B\kern-.05em{\sc i\kern-.025em b}\kern-.08emT\kern-.1667em\lower.7ex\hbox{E}\kern-.125emX}}
    
\begin{document}

%
\title[ASSED - Adaptive Social Sensor Event Detection]{ASSED - A Framework for Identifying Physical Events through Adaptive Social Sensor Data Filtering}

%
\author{Abhijit Suprem}

\email{asuprem@gatech.edu}

\affiliation{%
  \institution{Georgia Institute of Technology}
  \city{Atlanta}
  \state{GA}
}

\author{Calton Pu}
\email{calton.pu@cc.gatech.edu}
\affiliation{%
	\institution{Georgia Institute of Technology}
	\city{Atlanta}
	\state{GA}
}

%
\renewcommand{\shortauthors}{Abhijit Suprem and Calton Pu}

%
\begin{abstract}
	
Physical event detection has long been the domain of static event processors operating on numeric sensor data. This works well for large scale strong-signal events such as hurricanes, and important classes of events such as earthquakes. However, for a variety of domains there is insufficient sensor coverage, e.g., landslides, wildfires, and flooding. Social networks have provided massive volume of data from billions of users, but data from these generic social sensors contain much more noise than physical sensors. One of the most difficult challenges presented by social sensors is \textit{concept drift}, where the terms associated with a phenomenon evolve and change over time, rendering static machine learning (ML) classifiers less effective. To address this problem, we develop the ASSED (Adaptive Social Sensor Event Detection) framework with an ML-based event processing engine and show how it can perform simple and complex physical event detection on strong- \textit{and} weak-signal with low-latency, high scalability, and accurate coverage. Specifically, ASSED is a framework to support continuous filter generation and updates with machine learning using streaming data from high-confidence sources (physical and annotated sensors) and social networks. We build ASSED to support procedures for integrating high-confidence sources into social sensor event detection to generate high-quality filters and to perform dynamic filter selection by tracking its own performance. We demonstrate ASSED capabilities through a landslide detection application that detects almost 350\% more landslides compared to static approaches. More importantly, ASSED automates the handling of concept drift: four years after initial data collection and classifier training, ASSED achieves event detection accuracy of 0.988 (without expert manual intervention), compared to 0.762 for static approaches.

\end{abstract}

%
%


%

%

%

\maketitle

\section{Introduction}
\label{sec:intro}
Increasing volume of web and social media data combined with proliferation of Internet connectivity has paved the way for \textit{social sensors}, which are composed of social media and web streams, to have global coverage. These social sensors, including Twitter streams, Facebook posts, blog posts, and other web-data generated by humans, provide access to massive amounts of raw data that can be mined to obtain information for a variety of domains ~\cite{social_controversy, dis_mgmt_hagen, dis_mgmt_wang, dis_mgmt_imran}, including physical event detection. This entails building \textit{social source event detection} frameworks to support development of applications that ingest social sensor data and perform simple and complex physical event detection. A key challenge in this area (complex physical event detection) lies in the distinction between numeric data – a staple of complex event processing – and social sensor data; the latter is a primarily short text continuous stream that is difficult to represent with human-readable numeric features. Moreover, social network text data is \textit{live data} characterized by a language model that incorporates not only user demographics and geography, but also time ~\cite{twt_lang_model}, causing terms and features associated with a physical events to change with time. For example, text data about landslide events change in their data distribution in October and November in the United States because \textit{landslide} is often used to refer to election results more than the actual disaster \textit{landslide}. We refer to the terms, features, and signals associated with a physical event at any time as \textit{event characteristics}.

We differentiate physical event detection on social sensors from trend analysis: (i) we design our framework to accommodate strong- and weak-signal events, while trend analysis is usable on only the former and (ii) trend analysis assumes data is generated by a known stochastic model; conversely, as we show in Section ~\ref{sec:socialSensor}, social media text data's language model incorporates unknowable features. 

Therefore, social sensor text data presents a significant challenge for traditional ML classifiers and complex event processing systems, which operate in an offline or stationary setting where event characteristics do not change with time. Social source event detection systems that do not incorporate adaptation to evolving text streams face performance degradation over time; a representative example is Google Flu Trends (GFT). GFT was originally created to complement the CDC’s flu tracking efforts by identifying seasonal trends in the flu season ~\cite{gft_fail_c}. Failure to account for seasonal changes in event characteristics led to increasing errors over the years, and by 2013, GFT missed the trends by 140\%. This error has been attributed to exclusion of new data from CDC, changes in the underlying search data distribution itself, and cyclical data artifacts ~\cite{gft_fail_c,gft_fail_a, gft_fail_b}. We address these pitfalls by building adaptive event detection applications that continuously generate high-quality event detection filters and update existing filters. 

Specifically, we have the following contributions:
\begin{enumerate}
	\item We present ASSED – an Adaptive Social Sensor Event Detection framework to perform physical event detection with social sensors using an ML-based event processing engine.  Through ASSED, we show a practical framework for \textit{live data} processing, metadata extraction, and integration with high-confidence sources. We also present continuous automated machine learning filter generation, update, and retrieval procedures that work in concert to adapt to concept drift in live data.  
	\item We develop a Heterogenous Data Integration (HDI) process within ASSED to combine high-confidence sources with social-sensor data to automatically create labelled data for the ML-based event processing engine. High-confidence source are trustworthy, but have high latency and are scarce. Conversely, social-media sources have low confidence and are extremely noisy, but are abundant, and have global coverage with very low latency.  HDI integrates events from high-confidence sources into social sensor event detection to reduce noise, increase context and trustworthiness, and continuously create training data for ML classifiers. HDI allows ASSED to adapt to concept drift in live data and maintain high accuracy without manual intervention, making ASSED highly scalable.
	\item We believe ASSED is a useful framework for creating various physical event detector applications that would need social media sources. We demonstrate ASSED’s efficacy with an experimental evaluation application for disaster detection, specifically landslide detection, that outperforms static approaches. Our evaluation application, called LITMUS-ASSED, detects more than 3 times as many disasters as static approaches while maintaining high detection accuracy over several time windows.
	
\end{enumerate}

We focus on landslide detection because they are a class of disaster that do not have dedicated physical sensors or reputable trackers (in contrast to tsunamis and earthquakes), even though they cause significant monetary and human losses. Landslides are a weak-signal disaster with a lot of noise in social media streams; the use of the word \textit{landslide} to reference disasters (as opposed to election landslides or other usages) is dwarfed by posts about irrelevant topics that also use the word \textit{landslide}. 

We compare our application to LITMUS-original ~\cite{litmus_a}, which is a static landslide detector. We show significant detection improvements by identifying almost 350\% more landslide events through our application built on ASSED compared to static approaches. Applications built on ASSED are highly scalable to large numbers of physical event types. ASSED’s latency is dominated only by sensor delays, which is the timestamp difference between physical event occurrence and social media posts about the event. These delays are out of scope as they reside with the data sources, not with ASSED. To evaluate ASSED as a framework, we borrow accuracy metrics from machine learning such as precision, recall, and f-score. From 2014 through 2018, we show f-score of 0.988 in LITMUS-ASSED compared to f-score of 0.762 in LITMUS-original.

LITMUS-ASSED is currently running at  \texttt{\url{https://grait-dm.gatech.edu/demo-multi-source-integration/}} on landslide detection using multiple sources (USGS Earthquakes, TRMM Rainfall, and social media streams).

The rest of the paper is organized as follows: Section~\ref{sec:related} covers related work, Section~\ref{sec:ssed} introduces social source event processing, Section~\ref{sec:dataflow} introduces the ASSED dataflow, Section~\ref{sec:mlflowHDI} covers ASSED's ML classifiers and concept drift resiliency, Section~\ref{sec:application} evaluates the landslide detection application, and Section~\ref{sec:conc} concludes with future directions.

\section{Related Work}
\label{sec:related}
\subsection{Physical Event Detection}
Earthquake detection using social streams was proposed in ~\cite{dis_mgmt_sakaki}. This approach treats each user as a sensor and uses statistical filtering techniques to perform physical event detection. There have also been attempts to develop physical event detectors for other types of disasters, including flooding ~\cite{flood_detection}, flu ~\cite{gft_fail_c, disease_mgmt_wakamiya}, infectious diseases ~\cite{disease_mgmt_hirose}, and landslides ~\cite{r_esa, litmus_a}. More recently, burst detection for earthquake detection ~\cite{bursty_dis_mgmt} has been used to take advantage of the strong-signal nature of earthquakes. Strong-signal refers to size of event in data source; earthquakes and hurricanes are examples of strong signals as they usually have tens to hundreds of thousands of tweets. Conversely, landslides in ASSED have at most 5-10 tweets associated with them, making them weak-signal events.

Most approaches focus on specific large-scale disasters or health crises, such as earthquakes, hurricanes ~\cite{dis_mgmt_thom}, and influenza. We create ASSED to be a general-purpose physical event detector that can handle both large- and small-scale disasters, the latter having very small social and physical sensor footprint with significantly higher noise (weak-signal events). The existing approaches also assume static data, and as GFT has shown ~\cite{gft_fail_a, gft_fail_b}, such assumptions, though they can create accurate event detection in the short-term, degrade in the long term. Considering the volume of social-source data, manual tuning and updating of event detection filters and patterns can be prohibitively expensive.

\subsection{Traditional Mining and Static Event Processing}
The ubiquity of streaming data has increased work on adapting the static, offline setting to the dynamic setting characterized by changing, or \textit{drifting}, characteristics of events ~\cite{conc_drift_almeida, adversarial_drift, conc_drift_demello, conc_drift_active_shan, rejection}. There remain many assumptions of the static mining model in these adaptations, however  ~\cite{gama_drift_a}: (i) datasets are static/closed, (ii) immediate feedback is available, and (iii) direction and type of drift are known. CEP engines are unable to operate on detecting evolving events in dynamic environments and face scalability and performance issues in most approaches ~\cite{cep_uncertain_akila}.

\paragraph{\textsc{\textbf{Static/closed dataset}}}
Approaches described in ~\cite{gama_drift_a} assume streaming datasets are completely specified by feedback (which is also assumed to be given in real-time). Social sensor short-text data is difficult to characterize without labeled feedback, due in part to lexical diffusion ~\cite{lex_diff} and naturally evolving features of the data, such as geography, demographics, or time ~\cite{twt_lang_model}. 

\paragraph{\textsc{\textbf{Immediate feedback}}}
Most real-world streaming datasets are unstructured, schema-less web data. Real-time feedback is not possible, and delayed feedback is infeasible in web-scale applications. A single application operating on Twitter streams must consider over 500M tweets daily; manually labeling even 0.01\% of this data will require 20 workers each day to label tweets for 8 hours. Applications built on ASSED avoid relying on any human-generated feedback by inferring physical event detection filters and patterns from reputable sources such as physical sensors and news articles.

\paragraph{\textsc{\textbf{Known drift}}}
Most learning algorithms assume a single labeled dataset that is completely representative of all real-world data ~\cite{gama_drift_a, gama_drift_b}. Drift adaptive algorithms assume enough human-labeled data for each window is readily available, thereby assuming each data window is in the static, offline setting. ASSED considers data as dynamic with unknown, unbounded drift and does not require human-labeled training data for any window.

\subsection{Drift Adaptive ML Classifiers}
Current works in concept drift detection and adaptation use closed and non-evolving or synthetic data with known drift ~\cite{conc_drift_almeida, rejection, paired_learner, gama_drift_c}. \textbf{Windowing} is a common technique in drift correction. ~\cite{novelty_detection} uses a weighted k-Nearest Neighbor classifier is used to bin data into long-term or short-term memories. \cite{conc_drift_windows} considers multiple, nested windows to obtain feedback at different timescales. \textbf{Adaptive Random Forests }use an ensemble of weak decision trees combined with a drift detector to continuously prune and replace trees that degrade due to drift. Experiments conducted using weighted voting outperform simple majority voting \cite{arf_gomes}. \textbf{Knowledge Maximized Ensemble }(KME) \cite{kme} combines several drift detectors to identify cyclical, real, and gradual drift occurrences, updating models if either drift is detected or enough training data is collected for an update. Most strategies assume of availability of human feedback or expert-generated rules. We develop ASSED to be expert-independent in performing event processing on social sources. ASSED detects events and adapts to drift without manual intervention.

\section{Challenges of Event Detection from Social Sources}
\label{sec:ssed}
We introduce adaptive social source event detection as a new approach for detecting strong- and weak-signal physical events using social media streams. We distinguish social-sources from high-confidence sources; the latter consist of both physical sensors and reputable live data such as news articles. Strong-signal events in social and high-confidence data are the primary focus in traditional static event processing. We built ASSED to be a general purpose physical event detector in social sources that can support strong- \textit{and} weak- signals, and adapt to concept drift in live social data. This section elaborates on challenges in physical event detection from social sources, such as natural language processing (NLP), noise and concept drift, and detection of weak-signal events


\paragraph{\textsc{\textbf{NLP on Social Data}}}
\label{sec:socialSensor}
Social sensors consist of accounts and users from social-sources that provide text data along with various metadata. Metadata in social sensors consists of the multimedia content of a post, timestamp, location, links to other posts, and captions. NLP plays a key role in extracting usable information from social media text, and LITMUS-original provides several procedures for extracting location and analyzing text content for landslide detection. However, NLP is best used on long-text data; the text content of social-source posts is considered short-text data since the text does not provide enough words for most learning classifiers or NLP techniques ~\cite{short_text_sriram}. Our framework improves and augments the NLP procedures in LITMUS-original to improve coverage and reduce loss of data.

\paragraph{\textsc{\textbf{Noise and Concept Drift}}}
ASSED uses social sources in context of live data with concept drift. We use the language models introduced in ~\cite{twt_lang_model, lex_diff} to represent social data; our language model takes into account user demographics, post location, and timestamp:

\begin{equation}
\label{eq:textModel}
w_i = \operatorname*{arg\,max}_{w_i} \, P(w_i|{w_{i-1},u_{sp},l_{sp}, t_{sp}})
\end{equation}

where $P(w_i\mid\star)$ is the probability the next word in the sequence is $w_i$, $u_{sp}$ is the user who created the post ~\cite{twt_info_cred}, $l_{sp}$ is the location of the social post, and $t_{sp}$ is the timestamp of the social post. The reliance on user characteristics and location of user adds noise to social media data ~\cite{conc_drift_costa}. The reliance on time as a factor in text content is a characteristic of drift – as time changes, the distribution of text also changes. ASSED deals with noise and drift by continuously tuning machine learning (ML) filters. ASSED performs this tuning by integrating high-confidence but scarce reputable data with low-confidence but abundant social data.

\begin{figure}[h]
	\centering
	\includegraphics[width=\linewidth]{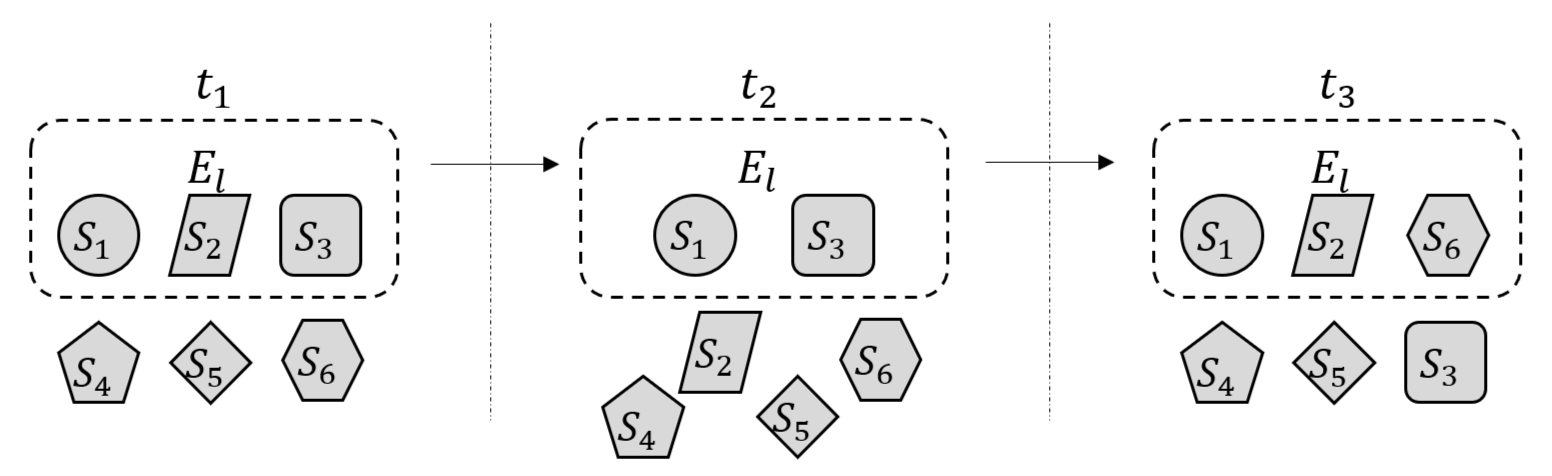}
	\caption{From time $t_1$ to $t_2$, the characteristics of event $E_l$ change. In $t_1$, $E_l$ is characterized by signals $S_1$,$S_2$, and $S_3$. However, in $t_2$, $E_l$ is characterized by signal $S_1$ and $S_3$ only.}
	\Description{Drifting event characteristics}
	\label{fig:driftingevents}
\end{figure}

\paragraph{\textsc{\textbf{Weak-signal Physical Event Detection}}}
ASSED is designed to support strong- and weak-signal events. We describe our model for physical events using our case study landslide detection application. 

Given the set of all events $\mathbf{E}$, we  denote our desired physical event as $E_{landslide}\in\mathbf{E}$. Each social post $P_i$ comprises of reporting on several events $E_a\in\mathbf{E}$. Each post is also independently composed of several signals $S_i$. A signal $S_i$ is analogous to a feature or column of data in machine learning. We thus define an event as a sum of signals:  

\begin{displaymath}
E_a = \sum_{k}a_i S_i
\end{displaymath}

where each $a_i$ is the strength of signal $S_i$ in event $E_a$. The evolving nature of short-text streaming data entails dynamic (or \textit{drifting}) $a_i$ over time, changing which signals are present in a specific event in any data window (Figure~\ref{fig:driftingevents}).

\begin{figure*}[t]
	\centering
	\includegraphics[width=\linewidth]{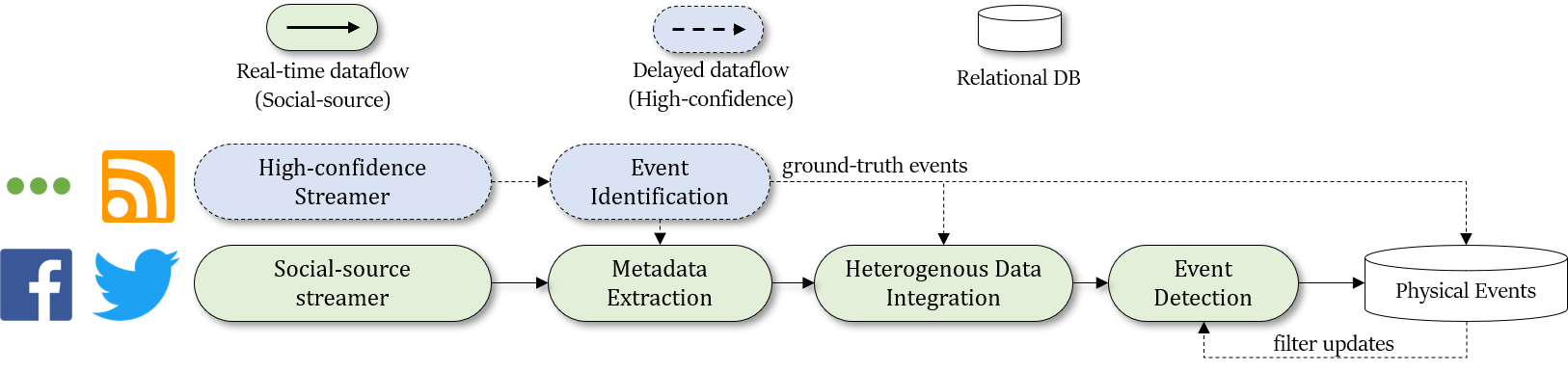}
	\caption{The ASSED dataflow combines live data from social sensors and high-confidence data from reputable sensor. An in-memory Redis-store is used to stage intermediate operations to pipeline physical event detection. ASSED combined high-confidence event processing (blue, upper) and social-source event processing (green, lower). ASSED remains adaptive by integrating high-confidence sensors with social sensors to continuously generate high-quality physical event detection filters.}
	\Description{ASSED Dataflow}
	\label{fig:assed_dataflow}
\end{figure*}

Strong-signal events have separable coefficients $a_i$ that can be clustered using unsupervised methods; trend analysis is a common technique for identifying large signal events. Since weak-signal events have small-valued coefficients with lower frequency compared to large signal events, trend analysis is not feasible. Human-generated rules are also impractical due to scalability issues ~\cite{cep_uncertain_akila}. ASSED is designed to support weak-signal event detection in addition to strong-signal events. Our framework uses machine learning combined with automatically generated training data to create filters for weak-signal events. ASSED avoids the requirement of human labelers by integrating high-confidence sources into the social-source dataflow to automatically generate ML training data for each data window.

\section{The ASSED Dataflow}
\label{sec:dataflow}
The ASSED dataflow is shown in Figure~\ref{fig:assed_dataflow}. ASSED performs physical event detection by integrating two sub-dataflows: \textbf{high-confidence dataflow} (blue, upper in Figure~\ref{fig:assed_dataflow}) and \textbf{social-source dataflow} (green, lower in Figure~\ref{fig:assed_dataflow}). The latency between a physical event’s occurrence and social sensor post about the event is significantly lower than latency with high-confidence sensor. This occurs because high-confidence sensors perform human annotations and confirm physical events, which are time-consuming. Social-source data is also abundant and has global coverage. However, it lacks the reputability of high-confidence sources. ASSED performs event detection by training machine learning filters to classify dense, fast social-source data as either relevant or irrelevant to landslides, with learning filters automatically updated using sparse, slow high-confidence data. Traditional approaches perform this under a static data assumption. ASSED’s novelty is in performing continuous updates by tuning its machine learning filters without any manual intervention to adapt to drift in social-source data. It accomplishes this by first, using the Heterogeneous Data Integration process to automatically match some social-source posts to high-confidence data for training and second, continuously updating machine learning filters using this data (covered in Section~\ref{sec:mlflowHDI}).

ASSED stores intermediate data for each process through a publish/subscribe interface. Our implementation uses a REDIS database (\textit{R\_Store}). Each process exports its outputs to \textit{R\_Store}. Decoupling intermediate processes from each other in the dataflow improves ASSED’s flexibility because each process can be developed, updated, and managed independently. A process exports each piece of data as a key-value pair to \textit{R\_Store}; the key is formatted using an \texttt{export-key} template. Each process registers its \texttt{export-key} with ASSED. Exported pairs are marked as \texttt{unprocessed} by ASSED. Processes also register an \texttt{import-key} template if they are obtaining \texttt{unprocessed} data from other processes. Processes that import a key-value pair inform ASSED after they have completed their task. Multiple processes can register the same \texttt{import-key} template. ASSED marks the corresponding key-value pair in \textit{R\_Store} as \texttt{processed} once all processes with matching \texttt{import-key} templates have completed their tasks on the pair. Decoupling also preserves intermediate data during process crashes; in the case of process failure, intermediate data in \textit{R\_Store} remains in an \texttt{unprocessed} state until ASSED restarts the failed process. \texttt{Processed} key-value pairs are scheduled for deletion by ASSED. 

\subsection{ASSED Case Study - Landslide Detection}
We develop a real-time critical physical event detection application for landslide detection on ASSED called LITMUS-ASSED. LITMUS-ASSED is an improvement on LITMUS-original ~\cite{r_esa,litmus_a}, which downloads streaming short-text (e.g. social media posts) to perform landslide detection using static classifiers operating under a closed dataset assumption. Landslide detection using social sensors is challenging: it is an open-ended live data with noise and the signal drift in social sensor data is unpredictable compared to synthetic and numerical data. Furthermore, keyword filters on streaming data are not enough for detection as \textit{landslide} has multiple meanings and can refer to a variety of topics, such as the disaster, elections, or the song "Landslide" by Fleetwood Mac. Moreover, landslides are a weak-signal event as the fraction of relevant signals for detection is significantly smaller than the fraction of irrelevant signals. 

However, performing landslide detection on social sensors is necessary: due to lack of coverage, lack of access, or absence of sensors, physical sensors are not enough to deliver dense, fast, global physical event detection. Landslides themselves cause several billions of dollars of damage, and fast event detection can be instrumental in limiting losses of life, reducing economic impact, and slowing damage progress. 

The rest of this section covers data Streamers, High-Confidence Event Identification, and Metadata Extraction. We cover Heterogenous Data Integration and ML Classifiers in Section~\ref{sec:mlflowHDI}.

\subsection{Social-source Streamers}
\label{sec:streamers}
ASSED’s architecture supports scalable streamers for high-confidence and social-source dataflows. Each streamer in an ASSED application must be deployed as either a high-confidence streamer or social-source streamer script. Streamers write data to the \textit{R\_Store} using their registered export-key template (key details provided in Table~\ref{tab:ssepDesc}).
\begin{displaymath}
{ss:en:landslides:Twitter:\{tweet\_url\}:7656:1550244443}
\end{displaymath}
is an example of a streamer \texttt{export-key} for the \textit{R\_Store} in the social-source dataflow. We have omitted the exact \textit{URL} for privacy.

\begin{table}[h]
	\caption{\texttt{Export-key} attribute descriptions for social-source and high-confidence source streamers in our landslide detection application. For some attributes, both types of streamers store same value.}
	\label{tab:ssepDesc}
	\begin{tabular}{|p{2.7cm} |p{5cm}|}
		\hline
		\textbf{key attributes} & \textbf{Social-source description}\\& \textbf{High-confidence description}                                          \\ \hline
		\texttt{streamer}                       & `social-source' or `ss'\\&`reputable-source' or `rs'                                                           \\ \hline
		\texttt{lang}                           & Langs supported by ASSED (`en') \\& `num' for physical sensors \\ \hline
		\texttt{key}                        & Physical event (`landslides')                                           \\ \hline
		\texttt{src}                         & Social network (`Twitter') \\& High-confidence Source (`NOAA')                  \\ \hline
		\texttt{url}                            & URL of social post \\& `NULL' if it is a physical sensor                  \\ \hline
		\texttt{id }                      & Auto-incrementing numeric ID                                      \\ \hline
		\texttt{timestamp}            & Timestamp of commit to \textit{R\_Store}                                   \\ \hline
	\end{tabular}
\end{table}

ASSED provides streaming templates for several social sensors including Twitter and Facebook, with an extensible framework for integrating additional streaming sources. These streamers operate in a high-volume setting, processing several thousands of tweets per second. We define each social post as a tuple  

\begin{equation}
\label{eq:tuple}
P_i=(p_i,\mathbf{l}_i,t_i,\mathbf{hl}_i,u_i)s
\end{equation}

 where $p_i$ is the post content as a Unicode string, $\mathbf{l}_i$ is an array of named locations within the post, $t_i$ is timestamp of post, $\mathbf{hl}_i$ is an array of hyperlink content within the post, and $u_i$ is the user-id or screen name. This tuple is published to the \textit{R\_Store} under its corresponding key defined earlier. $l_i$ is often null-valued as most social sensor posts do not provide geolocation information. ASSED uses the Metadata Extraction process to populate any null-values in the $P_i$ tuple. Any trivial metadata extraction, such as extracting user name from a tweet object or Facebook JSON object, can be performed within the streamer.

\subsubsection{LITMUS-ASSED Implementation}
LITMUS-ASSED uses ASSED's social streamers with common disaster keywords to download data from social sources. LITMUS-ASSED uses Twitter and Facebook sources.
\begin{enumerate}
	\item \textbf{Twitter}: a keyword streamer is used to download tweets continuously from Twitter. Keywords include the words `landslide', `mudslide', and `rockslide' as well as their lemmas.
	\item \textbf{Facebook}: an off-the-shelf keyword streamer is used to download public Facebook posts. Existing web crawlers (e.g. Google Search API) are leveraged to improve retrieval efficiency.
\end{enumerate}

$p_i$ is the short-text content (tweet or Facebook post). $\mathbf{l}_i$ is usually null as less than 0.5\% of tweets and posts provide geotagged locations. $t_i$ is post timestamp. $\mathbf{hl}_i$ is list of hyperlinks is provided in downloaded object. $u_i$ is user name of originating post.

\subsection{High-confidence Streamers}
High-confidence sensors are dedicated physical, social, and web sensors providing annotated physical event information; such sensor data is highly structured and contains detailed event information, including geographical coordinates and event time. However, their publishing latency makes them unsuitable for fast physical event detection – most high-confidence sensors report on events after multi-person confirmation and have significant delays in providing this information to the public. Additionally, such sensors do not have global or granular coverage, in contrast to social sensors. A high-confidence streamer in ASSED transforms raw data from each source into the same tuple structure as social-source streamers in Equation~\ref{eq:tuple}. 

\subsubsection{LITMUS-ASSED Implementation}
LITMUS-ASSED uses a variety of physical sensors and high-confidence sources for the high-confidence dataflow. Each stream element is exported to \textit{R\_Store}, with stream objects following the tuple structure described in Equation~\ref{eq:tuple}. The export-key's \textit{source} attributes are human-readable shortenings of the source (`usgs', `noaa', or `news').

\begin{enumerate}
	\item \textbf{USGS rainfall/earthquake reporting (physical sensor):}: LITMUS-original relied on USGS landslide reports~\cite{litmus_a}. Since USGS no longer provides up-to-date landslide reports for such disasters, LITMUS-ASSED streams daily rainfall and earthquake data from USGS. $p_i$ contains numeric sensor values. $\mathbf{l}_i$ and $t_i$ are location and timestamp provided by sensors, respectively. $\mathbf{hl}_i$ and $u_i$ are not applicable.
	\item \textbf{NOAA landslide predictions (high-confidence source):} The National Oceanic and Atmospheric Administration (NOAA) provide landslide predictions in select locations. LITMUS-ASSED streams the predictions along with prediction probability to augment high-confidence physical event detection. $p_i$ contains landslide prediction probability provided by sources. $\mathbf{l}_i$ and $t_i$ are location and timestamp, respectively. $\mathbf{hl}_i$ and $u_i$ are not applicable.
	\item \textbf{News articles (high-confidence source):} Since physical sensors do not have global coverage, LITMUS-ASSED also streams news articles downloaded from various online RSS feeds and aggregators (Google News and Bing News APIs). $p_i$ contains article summary provided by RSS feeds and aggregators and article title. $\mathbf{l}_i$ is article location is available in metadata, or \textit{null} otherwise. $t_i$ article publish time. $\mathbf{hl}_i$ is article link provided by RSS feed or aggregator. $u_i$ is article source (BBC, Marietta Daily Journal, etc).
\end{enumerate}

\subsection{High-Confidence Event Identification}
\label{sec:hcevents}

\begin{figure}[h]
	\centering
	\includegraphics[width=\linewidth]{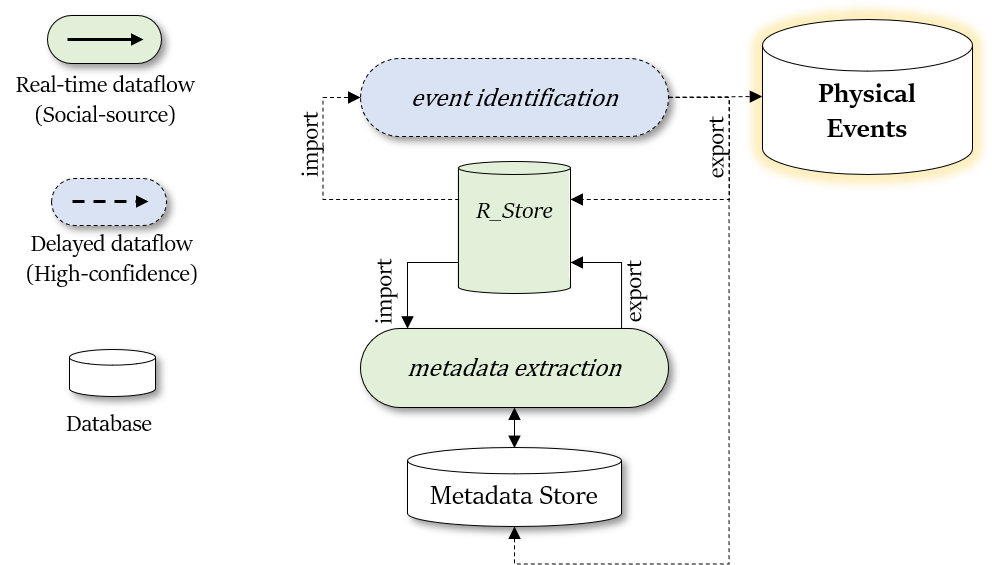}
	\caption{After data download, high-confidence dataflow performs event identification using physical sensors. Social-source dataflow continues with metadata extraction. Since social sensors have low-context data with missing attributes, metadata extraction attempts to fill the missing gaps. For both processes, information is retrieved from \textit{R\_Store} using their respective \texttt{import-key} templates (which match streamer \texttt{export-key} templates).}
	\Description{Event identification and metadata extraction}
	\label{fig:stage2}
\end{figure}

The high-confidence dataflow identifies ground-truth physical events from high-confidence streamer downloads. Since high-confidence sources provide annotations, extracting event information relies on using these annotations. Event metadata is shared with Social-source Metadata Extractors to augment their NLP procedures (Section~\ref{sec:metadata}). Figure~\ref{fig:stage2} describes this sharing process between Event Identification and Metadata Extraction through the Metadata Store.

For physical sensors, it is trivial to insert the physical event provided by the sensors into ASSED’s physical event relational database (\textit{PE\_RDB}). We adapt several implementation details from LITMUS-original. One of our contributions in this process is support for alternative high-confidence sensors such as news articles and reports. News articles provide topic tags that can be mined for an application’s event, and event reports (e.g. earthquake or large landslide report by USGS) provide detailed information about events, including locations, timestamps, event range, and event impact. Physical events in high-confidence dataflow are inserted to the \textit{PE\_RDB} with fields described in Table~\ref{tab:rdsSchema}. ASSED also provides a Metadata Store (\textit{M\_Store}) where physical event attributes can be shared with the social-source dataflow to augment Metadata extraction.

\begin{table}[h]
	\caption{\textit{PE\_RDB} schema for physical events detected in high-confidence sources (physical sensors and reputable sources)}
	\label{tab:rdsSchema}
	\begin{tabular}{|p{2.7cm} |p{5cm}|}
		\hline
		\textbf{Field} & \textbf{Description}                                                                                               \\ \hline
		\texttt{location}       & Latitude and Longitude. If named location is provided, we perform coordinates lookup using off-the-shelf maps APIs \\ \hline
		\texttt{event\_time}    & Event occurrence time as reported as reported by originating sensor                                                \\ \hline
		\texttt{sensor\_sourc}e & Tuple of physical sensor source name and URL (obtained from import-key attributes)                                 \\ \hline
		\texttt{event\_obj}     & Serialized high-confidence streamer object                                                                                    \\ \hline
	\end{tabular}
\end{table}

\subsubsection{LITMUS-ASSED Implementation}
We devise several event rules for physical event detection in the high-confidence dataflow in LITMUS-ASSED. We use each source and its data $D_s\rightarrow [D_{rain}$, $D_{NOAA}$, $D_{news}$, $D_{quake}$,$\cdots]$. $D_{NOAA}$ provides landslide predictions in select locations.

\begin{enumerate}[Event Rule 1.]
	\item If $D_{NOAA}.prediction>70\% \, \wedge $ ($D_{rain}$ shows rain within 3 days of $D_{NOAA}.time$) $\vee$ ($D_{quake}.mag>3$ within a day $\wedge$ $D_{rain}$ shows some rain within 3 days)
	\item If $D_{quake}.mag>6$ $\wedge$ $D_{rain}$ shows rain within 3 days $\wedge$ $D_{NOAA}.prediction>30\%$
	\item If $D_{quake}.mag>7$ $\wedge$ $D_{rain}$ shows rain within 3 days
	\item If $D_{news}$ has articles tagged with \textit{landslide} or \textit{mudslide} by reporting agency in any location
	
\end{enumerate}

For each event rule, if posts within 50km of a location trigger all conditions, we store a physical event within \textit{PE\_RDB}. All detected physical events’ locations are sent to \textit{M\_Store}. High-confidence posts that only partially matched event rules also have locations sent to \textit{M\_Store}, but are not stored in \textit{PE\_RDB}. \textit{M\_Store} exports have an expiration of 1 week.

\subsection{Social-source Metadata Extraction}
\label{sec:metadata}
Event detection applications have a variety of metadata requirements; ASSED provides extensibility in adding application-specific metadata extractors. We assume the following metadata scenarios:

\begin{enumerate}[1.]
	\item Applications requiring a single metadata extractor that operates on data from a social-source streamer
	
	\item Application requiring multiple metadata extractors operating sequentially.

	\item Applications requiring multiple metadata extractors, some of which operate on data from a social-source streamer; other extractors operate sequentially on prior extractors
	
\end{enumerate}
\vspace{.2cm}

Metadata extractors register their \texttt{import-key} and \texttt{export-key} template with ASSED. Scenarios 1 and 2 are handled by the decoupled dataflow, where each extractor will publish to the next process. To accommodate Scenario 3, ASSED marks a key-value pair $kv_i$ as \texttt{processed} only if \textit{all} processes with an \textit{import-key} template matching $kv_i$’s key also mark it as \texttt{processed}. Once all importers have accessed a key-value pair and performed their operations, it can be \texttt{processed} and deleted by ASSED. An example of a metadata \texttt{export-key} template is shown below, where \textit{ext\_name} is the name of the metadata extractor (e.g. \textit{Twitter\_location\_extractor}, or \textit{hyperlink\_extractor}).

 \begin{displaymath}
ext\_name:lang:key:source:url:post\_id:timestamp
 \end{displaymath}
 
 \paragraph{\textsc{\textbf{Augmentation with M\_Store}}}
 Social-source data has low-context and reputability, which hinders NLP metadata extractors. ASSED augments natural language extractors such as NER (Named Entity Recognition) with information shared by the high-confidence dataflow. We provide an example with location extraction. Since social posts have few words, location extraction is not accurate on the short-text and often misses locations provided in a post’s text content. Any location identified by a location extractor can be published to \textit{M\_Store} with an expiration timer (set by the extractor and is application-specific), after which it is deleted by ASSED. Other metadata extractors now have access to the location string in \textit{M\_Store} before its expiration to augment location extraction with e.g., substring match. Event locations identified in the high-confidence dataflow can also be used by social-source metadata extractors when available.

\subsubsection{LITMUS-ASSED Implementation}

\begin{figure}[h]
	\centering
	\includegraphics[width=\linewidth]{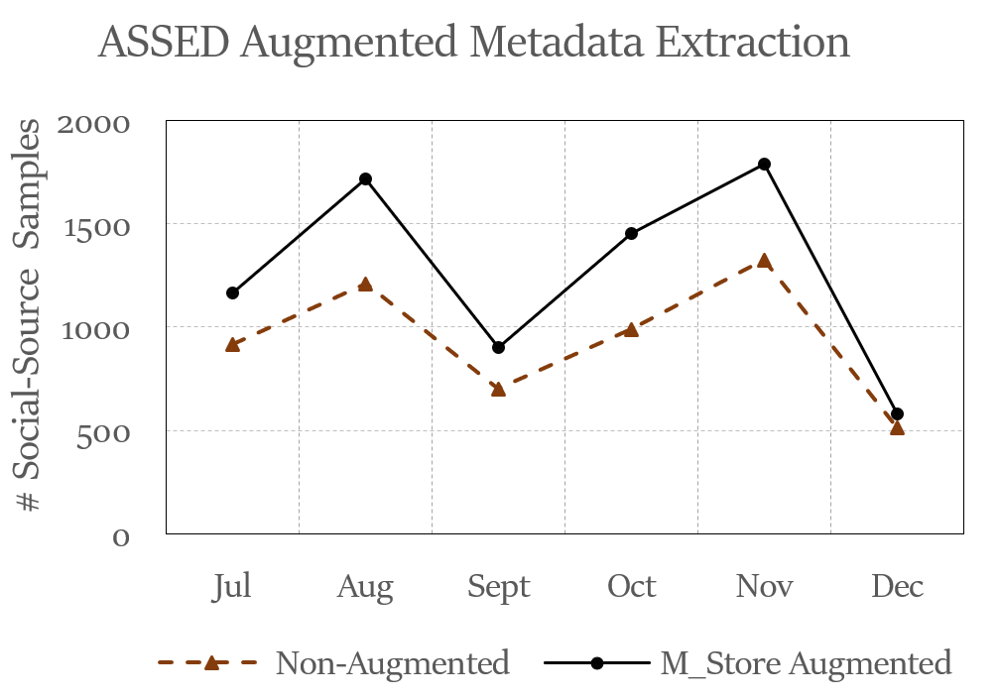}
	\caption{Lack of metadata for social-source posts leads to post deletion by ASSED. Augmenting metadata extraction with \textit{M\_Store} increases social-source data in LITMUS-ASSED in each monthly data window and reduced deletion of potentially relevant posts.}
	\Description{LITMUS-ASSED metadata augmentation}
	\label{fig:metadata}
\end{figure}

Posts imported from \textit{R\_Store} with null $\mathbf{l}_i$ are processed with a location extractor. Off-the-shelf NER~\cite{heterogenous_finkel} (Named Entity Recognition) is applied to social-source post content to identify locations. Since NER does not perform well on social posts, LITMUS-ASSED uses location strings from \textit{M\_Store} to perform substring match. Locations identified from NER are sent to \textit{M\_Store} with expiration of 2 days. Locations identified with substring match have expiration updated with additional 2 days if expiration is less than 2 days. We used 2 days as we observed most disasters detecteded in social-source dataflow had majority of posts within 2 days of first social sensor report. Using \textit{M\_Store} to augment NER increases social-source posts’ location extraction (Figure~\ref{fig:metadata}) and reduces discarding of potential useful data.

Post location(s) are converted to geographical coordinates and mapped to 2.5-min cell grids on the planet~\cite{litmus_a}. For each location ${L_1,L_2,\cdots,L_k }\in \mathbf{l}_i$ for a post, LITMUS-ASSED obtains coordinates using online map APIs. Coordinates are mapped to 2.5-min cell grids on the planet using approach in ~\cite{litmus_a}. Since a post can have multiple locations, we group locations by cell grid. Cells with the maximum number of coordinates are accepted as post location.

\section{ML-Based Event Detection}
\label{sec:mlflowHDI}

ASSED performs ML-based event detection on social-source data. Traditionally, ML-based event detection is performed under static data assumption. Since ASSED is designed for social-source data, the static data assumption is not valid (we provide evidence for drift in our application in Figure~\ref{fig:driftEvidence}). ASSED continuously updates filters and stores prior and current filters in a filter database (\textit{F\_Store}). ASSED's filter update is performed without human or expert input, allowing high level of scalability considering the large volume of social sensor data. 

Since ASSED is designed to be an \textit{adaptive} event detection framework that is resilient to concept drift, we introduce a \textit{Heterogenous Data Integration} (HDI) process in ASSED to confer this adaptivity to ML classifiers. Heterogenous Data Integration matches events from high-confidence dataflow into social-source dataflow with joins to improve social sensor event detection and perform continuous filter updates on event detection classifiers. We call this step \textit{heterogenous} because ASSED is matching social sensor data (short-text, hashtags) to high-confidence sensor data (reports, physical sensors, news articles, etc). In contrast to static event processors, ASSED continuously improves and updates its event detection filters to adapt to drifting event characteristics in social sensor data. While this can be accomplished manually with experts, it is not practical for social-source event processing due to the massive volume of data (as mentioned, labeling 0.01\% of Twitter data will require 20 experts each day to label tweets for 8 hours). 

\subsection{Heterogenous Data Integration (HDI)}
\label{sec:HDI}
We perform HDI for two reasons: (i) data integration allows ASSED to improve social-source event detection by using metadata from high-confidence sources to expand metadata filters after extraction; and (ii) social posts that match high-confidence physical events can be used to continuously update and tune social sensor event detectors to prevent deterioration like GFT.  Joins are performed on event attributes using (a) string schema matching, (b) string similarity join, and (c) natural joins between key-value pairs in \textit{R\_Store} and physical events in \textit{PE\_RDB} detected in high-confidence dataflow.

\begin{enumerate}[(a)]
\item \textbf{Schema Matching}:
Since application developers are given control of key-template specifications for \textit{M\_Store}, they can also specify schema mappings to ASSED for data integration. It is then trivial to perform string similarity or natural joins on the \textit{R\_Store} and \textit{PE\_RDB} values.

\item \textbf{String Similarity Join}:
ASSED performs string similarity join using off-the-shelf similarity functions such as Jaccard and Levenshtein matching with a join predicate:

\begin{displaymath}
string\_sim(key\_attr,rsep\_field\_val)>threshold
\end{displaymath}

\item \textbf{Natural Joins}:
With access to schemas, natural joins are performed with a SQL query on \textit{PE\_RDB}.
\end{enumerate}

Joining on all physical events is impractical. A solution is variable data windows for the join range that can be user- or data- specified:

\paragraph{\textsc{\textbf{User-specified windows}}}

Application developers can pre-define a time range $t_0$ to select events from \textit{PE\_RDB}. Domain-experts can contribute their knowledge towards specification of these windows for each event. During integration at $t_i$, ASSED then selects all physical events within $[t_i-t_0,t_i]$ for joins.

\paragraph{\textsc{\textbf{Data-specified windows}}}

When new physical events are available in \textit{PE\_RDB}, ASSED can use the join predicate on social-source posts with fuzzy detection. For each physical event, ASSED finds all social-source events that have the same metadata (location, timestamp). Social-source posts within a threshold distance of \textit{PE\_RDB} events and within a threshold time of \textit{PE\_RDB} events are selected for joins.

Only a fraction of social-source posts are matched to events detected in high-confidence dataflow (we show an example with our landslide detection application in Figure~\ref{fig:sspLabel}). High-confidence-sources have large publish latency as they need humans to perform confirmation; as such it is not uncommon for high-confidence-sources to report an event several days after event occurrence. In contrast, social-sources report an event within a few hours at most in remote areas, and within minutes in urban regions. High-confidence-sources also do not have global coverage compared to social sensors, which are spread around the world. 

\subsection{LITMUS-ASSED Implementation}
\begin{figure}[h]
	\centering
	\includegraphics[width=\linewidth]{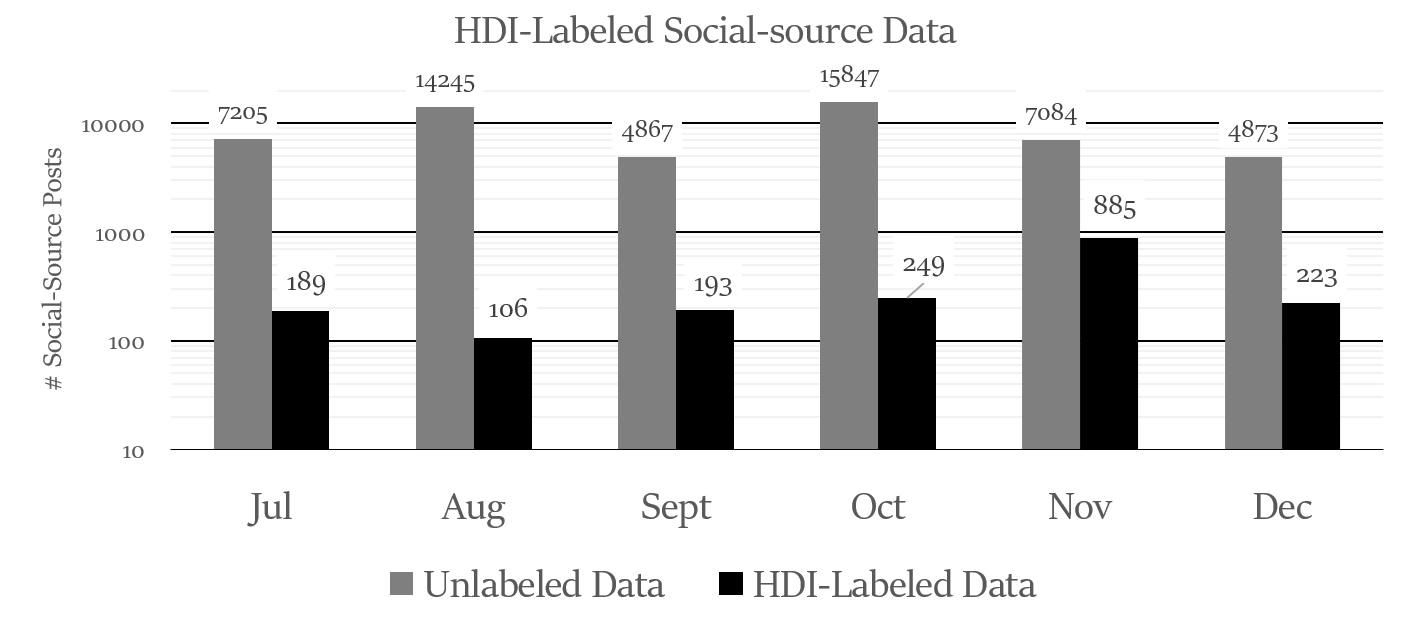}
	\caption{Labeled and unlabeled social-source dataflow posts (y-axis is log-scale). On average, less than 5\% of each window’s samples can be labeled.}
	\Description{Labeled and unlabeled}
	\label{fig:sspLabel}
\end{figure}

LITMUS-ASSED uses user-specified $\pm3$ days as the time filter for joins. Once an event is detected in the high-confidence dataflow, it is sent to \textit{PE\_RDB} and to the HDI process. HDI then retrieves all social-source posts from \textit{R\_Store} within 3 days prior (i.e. $-3$ days) to the physical event using its \texttt{import-key}. HDI then keeps the physical event in memory for 3 additional days (i.e. $+3$ days) to match future social-source posts in \textit{R\_Store}. ASSED’s string similarity and schema matching joins are performed on the physical events and social-source posts collect by HDI. Fuzzy matching on location and timestamp is used to reduce join time. Location matching is achieved with the 2.5-min cell grid superposition using strong supervision; intuitively, social posts that have landslide keywords and are made within the similar time at similar places as a landslide identified through a news source are very likely relevant to a landslide. This is in contrast to weak-supervision [40] as our supervisory labeling is domain-specific. Social-source posts that can be mapped to physical events from high-confidence dataflow by ASSED are labeled as training data (described in Section~\ref{sec:mlflow}). Posts that could not be labeled are used as prediction data for event processing with ML. Only a fraction ($<5\%$) of social-source posts can be labeled in HDI process (Figure~\ref{fig:sspLabel}, y-axis is log-scale).

\subsection{ML-based Event Detection}
\label{sec:mlflow}

\begin{figure*}[t]
	\centering
	\includegraphics[width=\linewidth]{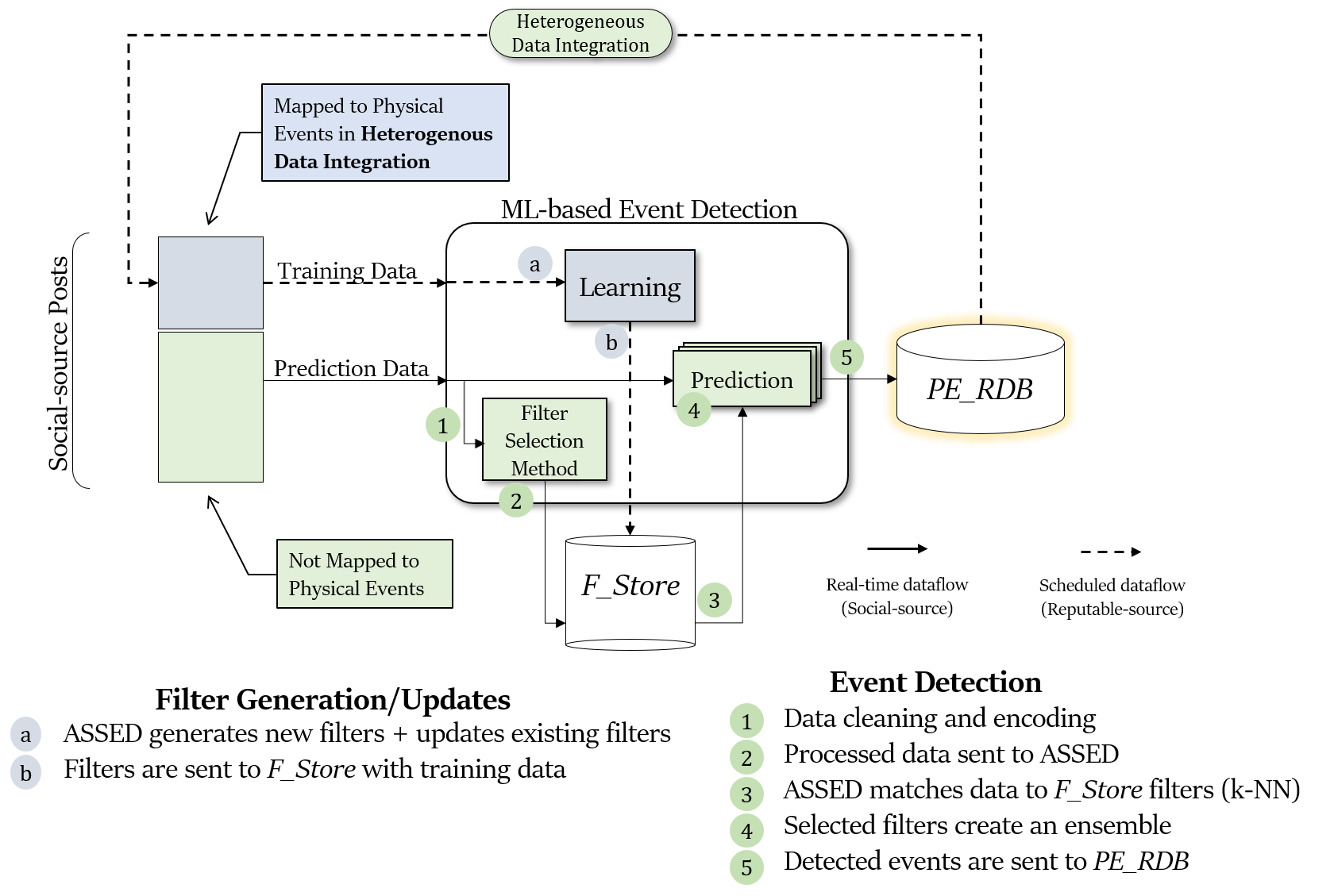}
	\caption{Event detection with ML and filter generation/updates in ASSED. Filter generation and updates are scheduled either by humans or based on drift-detection mechanisms as covered in Section~\ref{sec:generation} and ~\ref{sec:updates}. Each filter is stored with its training data. During event detection, new data is matched to its corresponding filter (i.e. current filter for User-specified windows, or closest training data using clustering for Detector-specified or Hybrid windows). The data is classified and relevant samples are sent to the Physical Events database.}
	\Description{ML Event detection}
	\label{fig:mlDataflow}
\end{figure*}

ASSED uses machine learning classifiers to perform event detection. Filter generation and updates (training) are performed with the subset of social-source posts from HDI that were mapped to physical events ($<5\%$ of social posts). Event detection (prediction) is performed on the remaining social-source posts that could not be mapped and are unlabeled. The process is described in Figure~\ref{fig:mlDataflow}. 

\subsection{Filter generation}
\label{sec:generation}

ASSED generates filters according to an application's update schedule. We first describe the filter generation/update procedure.

A data window is the set of social-source posts between the previous filter update and current time. At the end of a data window, all social posts that were mapped to physical events during HDI are labeled as \texttt{relevant} for the ML classifier. Rejected posts from social-source streamers are used as \texttt{irrelevant} samples. During initial configuration of an ASSED application, domain experts can also manually label some samples (this is a one-time cost, in contrast to common machine learning update approaches where manual labeling is required in all data windows). The labeled samples are used to train new ML classifier filters. If filters already exist, they are updated with the new data. Both new and updated filters are saved to \textit{F\_Store} as key-value pairs, with the filters as values and \textit{training data characteristics} as the key. Currently ASSED stores the entire training data along with training timestamp as the key. Filters are retrieved using either timestamp lookup (if searching for most recent window) or nearest neighbor search (if searching for most similar training data).

\subsection{Filter update schedule}
\label{sec:updates}
ASSED supports three types of filter update schedules: \textit{User-specified}, \textit{Detector-specified}, and \textit{Hybrid}, described below. These schedules allow for continuous filter generation and updates to adapt to concept drift.

\paragraph{\textsc{\textbf{User-specified}}}
Application developers or domain experts determine a predefined update schedule (monthly, weekly, etc). When an update is triggered, ASSED follows procedures in Section~\ref{sec:generation}.

\paragraph{\textsc{\textbf{Detector-specified}}}
Some machine learning classifiers provide confidence values with their predictions. In the case of SVMs, distance from hyperplane ($W^T x+b$) constitutes a measure classification strength. For linear classifiers (including SVMs), increased density of samples closer to decision boundary over time indicate signal drift ~\cite{md3}. Neural networks with softmax output layer provide class probabilities. Burst detection on increasing fraction of lower probability labels also indicate drift ~\cite{conc_drift_almeida}. If drift time exceeds a threshold, the event detector request filter generation and update.

\paragraph{\textsc{\textbf{Hybrid}}}
The Hybrid schedule combines \textit{User-} and \textit{Detector-} specified approaches. Users specify a schedule using domain knowledge. LITMUS-ASSED uses a monthly schedule. ASSED also track's each filter's classifications; when drift is detected with margin density ~\cite{md3}, a new filter is created using labeled data from the current data window. The existing filter is copied. The copy of the filter is updated with labeled data from the current window. This allows ASSED to keep incremental knowledge as filters in staggered windows.

\subsection{Classifier Selection and Weighting}
ASSED’s machine learning event detectors use multiple classifiers with voted majority to perform predictions, as an ensemble of classifiers is known to perform better than a lone classifier. ASSED allows a variety of weighting schemes for classifier ensembles:

\paragraph{\textsc{\textbf{Unweighted average}}} 

 Class labels predicted by each classifier in the ensemble (\texttt{0} for irrelevant and \texttt{1} for relevant physical event) are summed and averaged. $Score\geq0.5$ indicates majority of classifiers consider the input post as relevant to the physical event.
 
\paragraph{\textsc{\textbf{Weighted average}}} 

Classifiers can be weighted by domain experts based on which algorithm they implement. Weak classifiers (random forests) would be given lower weights than better classifiers (SVMs). Deep learners using convolutional networks or neural networks would get larger weights than statistical or linear classifiers (such as Logistic Regression).

\paragraph{\textsc{\textbf{Model-weighted}}}

 ASSED can determine classifier weights using their prior performance using Eq.~\ref{eq:mweights}. For a given dataset with $n$ classifiers, $C_i$ is classifier $i$, $w_{C_i}$ is the weight of classifier $i$, and $f_{C_i}$ is the validation accuracy (using \textit{f-score}) of $C_i$ on its testing data. Each classifier's weight is proportional to its performance.
	
	\begin{equation}
	\label{eq:mweights}
	w_{C_i^t} = \frac{f_{C_i}}{\sum_a^n f_{C_a}}
	\end{equation}

We support the common machine learning frameworks \textit{sklearn}, \textit{keras}, and \textit{tensorflow}, and are planning to integrate additional frameworks into ASSED.

\subsection{ML Classifiers in LITMUS-ASSED}

LITMUS-ASSED’s data is in the form of social posts, with short text. The text must be processed into a numeric format for classifiers in social-source dataflow. ASSED includes support for word2vec, GloVe, and Explicit Semantic Analysis (ESA) ~\cite{esa_main} encodings. We examined the performance of word2vec and ESA in LITMUS-ASSED in a separate machine learning experiment to identify encodings that are more resilient to signal drift. We found \textit{w2v} performed the best, and use it for LITMUS-ASSED. LITMUS-original uses ESA, which we found to be least resilient.

\subsubsection{Signal Drift}
\begin{figure}[h]
	\centering
	\includegraphics[width=.8\linewidth]{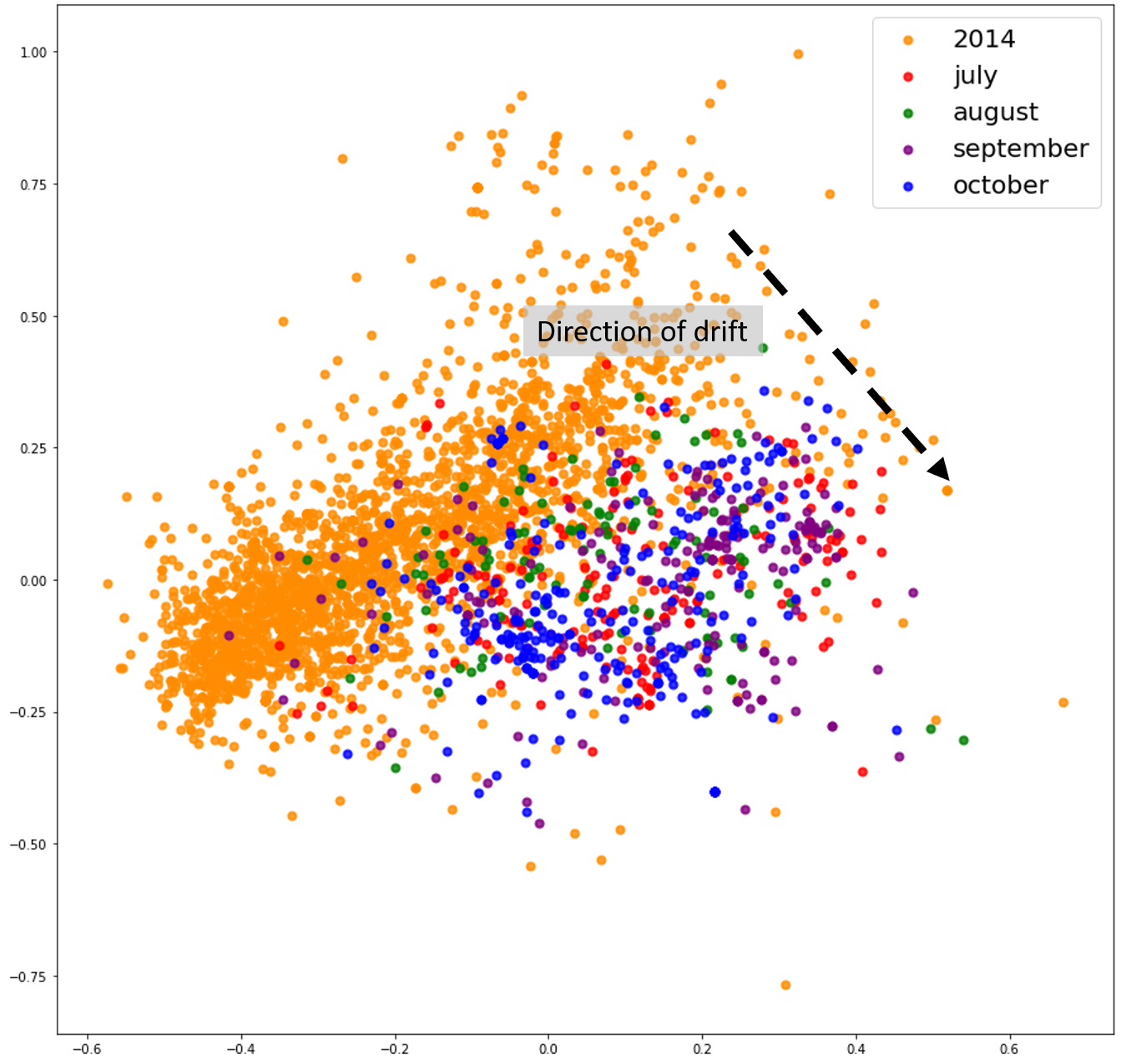}
	\caption{Signal drift changes event characteristics. We measure drift with changes in social posts labeled as relevant to landslides in LITMUS-ASSED. Axes are raw principal component scores normalized to  $[-1,1]$.}
	\Description{Evidence of drift in social-source data}
	\label{fig:driftEvidence}
\end{figure}

We first demonstrate the need for ASSED’s adaptive event detection with evidence of signal drift in our social sensor data. Each HDI-labeled social-source post’s text is converted to a high-dimensional, numeric representation using word2vec [14]. The post vector's dimensionality reduced with Principal Component Analysis; we train a PCA-reducer in \textit{sklearn} on the 2014 data to identify most important composite features in 2 dimensions (for visualization). The reducer is used to transform all data windows (2014, July-2018, August-2018, etc) and reduce them from 300-dimensions (the default for word2vec) to 2-dimensions. Changes in data window centroids indicate drift in Figure~\ref{fig:driftEvidence}. The axes correspond to raw principal component values. Our real-world live data will continue to evolve over time. 

\begin{figure}
	\centering
	\includegraphics[width=\linewidth]{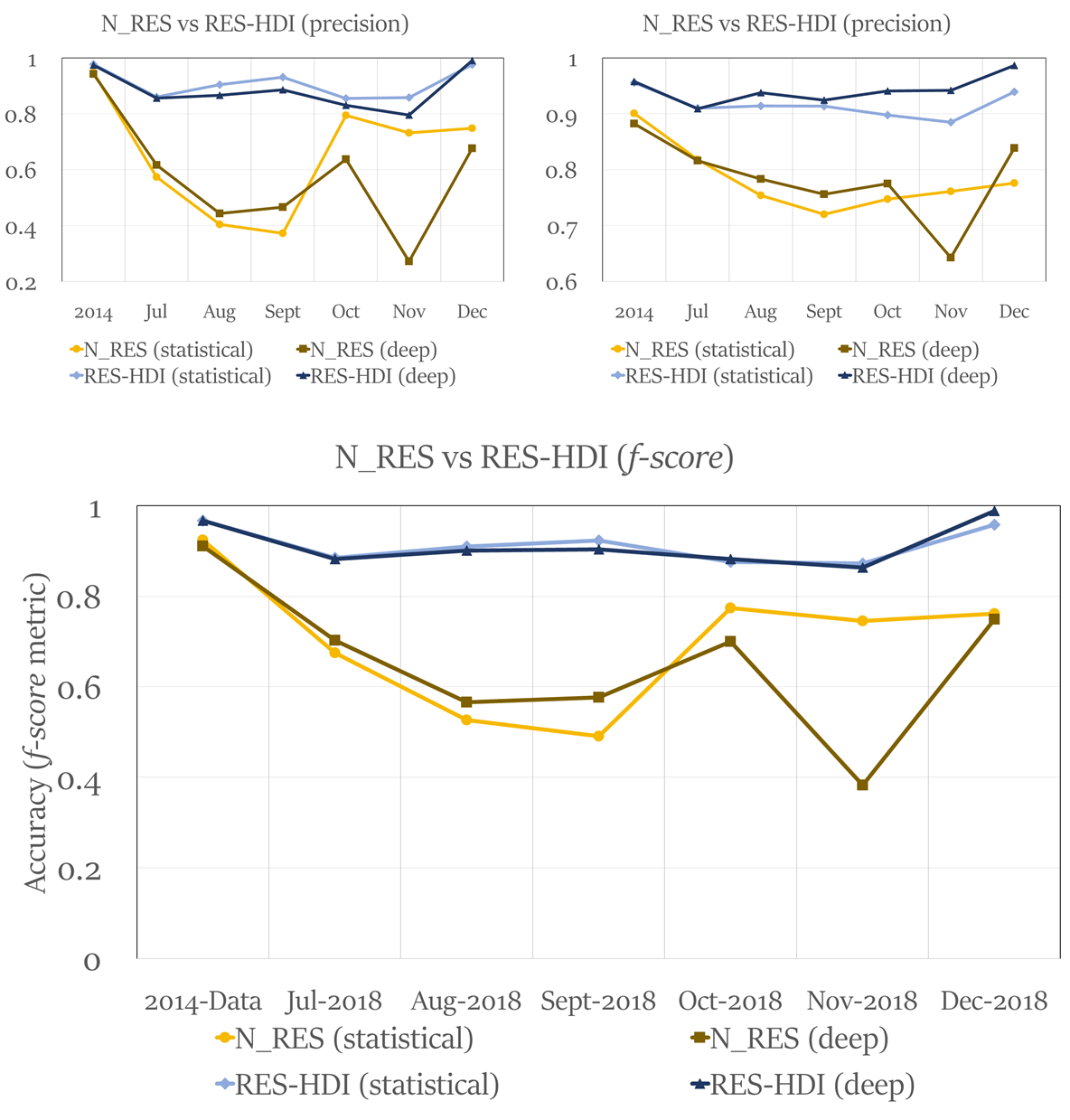}
	\caption{Both \texttt{RES-HDI (deep)} and \texttt{RES-HDI (statistical)} significantly outperform \texttt{N\_RES} counterparts. We show precision (top left) and recall (top right); \texttt{RES-HDI} has lower false positives (higher precision) and lower false negatives (higher recall) than \texttt{N\_RES}. The f-score (bottom) indicates \texttt{RES-HDI} has far lower variance in performance than \texttt{N\_RES}.}
	\Description{results}
	\label{fig:results}
\end{figure}

We also show performance decay of static classifiers during evaluation (\texttt{N\_RES} in Figure~\ref{fig:results}).

%
%

\subsubsection{Event detection approaches}
ASSED’s ML-based event detection approach integrates physical events into social-source event detection using HDI. We evaluate this integration by comparing physical event detection accuracy on social-source posts to traditionally static event processing, which perform event detection without HDI integration. We call the HDI-augmented approach RES-HDI, or Resilient-HDI, because it is concept drift resilient. 


The RES-HDI ensemble classifiers are weighted using ASSED’s model-weighted scheme in Equation~\ref{eq:mweights}.
  
Each approach is separated into \texttt{deep} and \texttt{statistical} versions based on classifier composition (deep learners for the former and statistical learners for the latter) as \texttt{N\_RES/RES (deep)} or \texttt{N\_RES/RES (statistical)}, respectively. Each approach is built as an ensemble of multiple machine learning classifiers.

\subsubsection{LITMUS-ASSED accuracy evaluation}
We show performance of each approach over several data windows in Figure~\ref{fig:results}. RES-HDI under both statistical and deep learners maintains high accuracy during the drift visualized in Figure~\ref{fig:driftEvidence}. \texttt{N\_RES} approaches have higher variance in performance and have deteriorating accuracy. HDI confers resiliency to ASSED’s filters, and it has a significant impact on event detection. By December 2018, \texttt{RES-HDI} with deep learning classifiers has \textit{f-score} of 0.988 while the static approaches have f-scores 0.762 and 0.7493 for statistical and deep learning, respectively. RES-HDI has a variance of 0.0015 versus \texttt{N\_RES}’s variance of 0.021, which is an order of magnitude greater.

\section{Landslide Event Detection with LITMUS-ASSED}
\label{sec:application}
\begin{figure}[h]
	\centering
	
	\includegraphics[width=\linewidth]{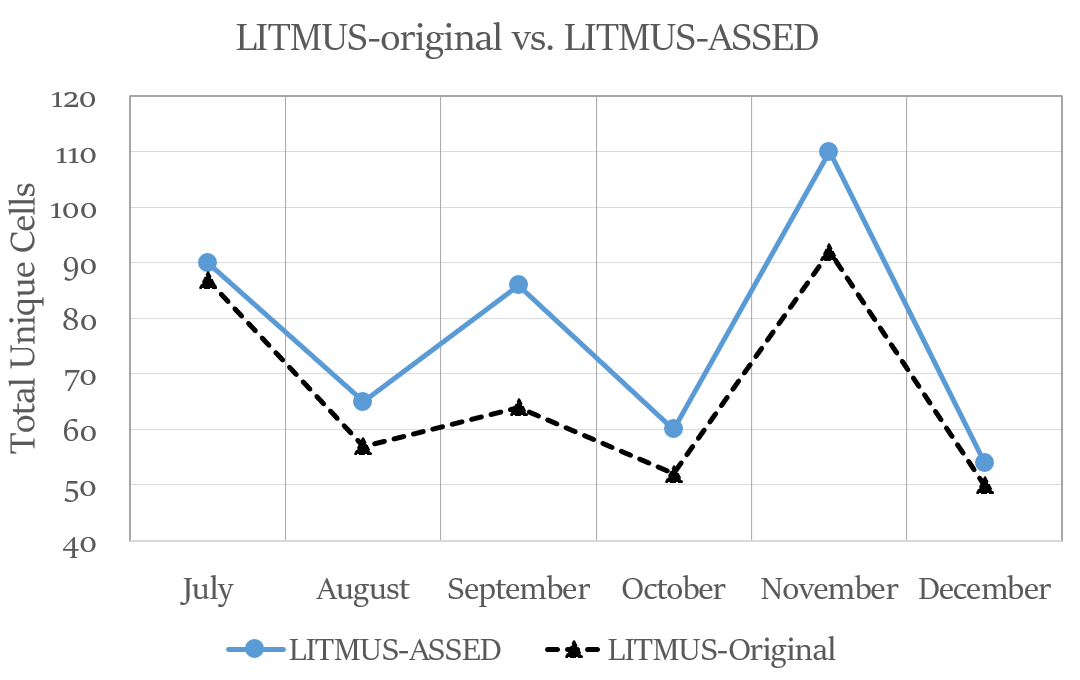}
	\caption{Additional total unique cells under ASSED compared to LITMUS-original. Multiple events can be mapped to a single cell. All events detected by LITMUS-original are also detected by LITMUS-ASSED.}
	\Description{Total Unique Cells}
	\label{fig:compareA}
\end{figure}

\begin{figure}[h]
	\centering
	
	\includegraphics[width=\linewidth]{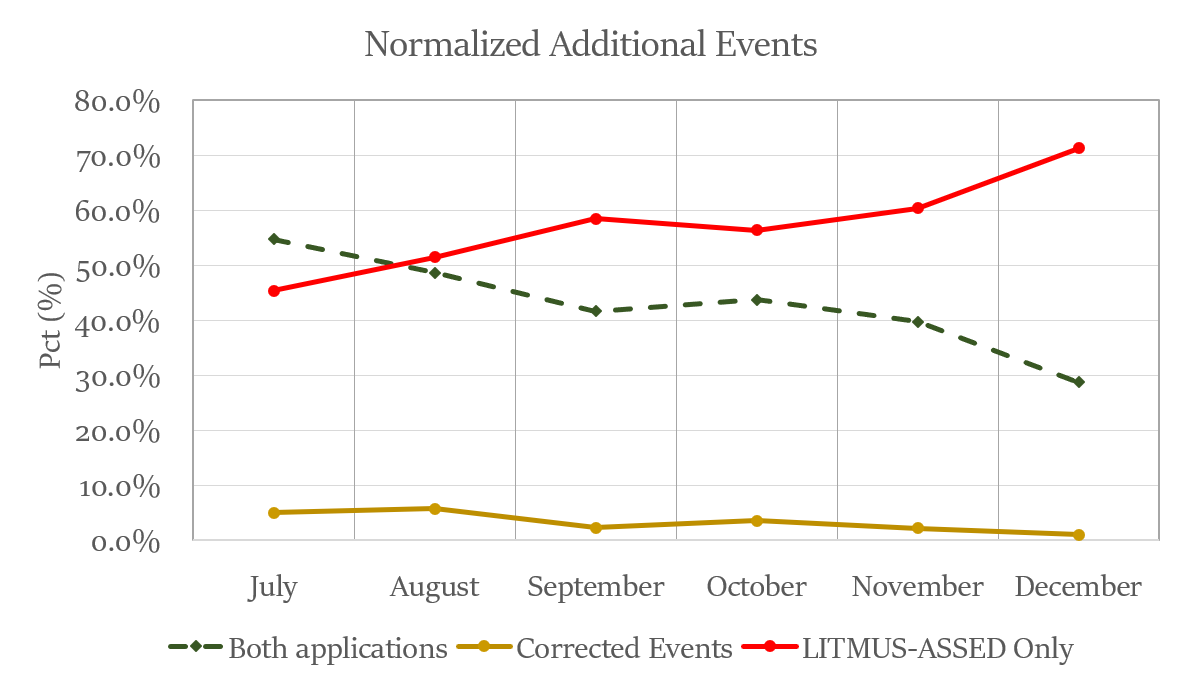}
	\caption{We compare the fraction of events detected only in LITMUS-ASSED and fraction of events detected by both approaches. LITMUS-ASSED detects several additional events that LITMUS-original fails to identify. LITMUS-ASSED also corrects false-positives made by LITMUS-original (\texttt{Corrected Events}).}
	\Description{Additional Events}
	\label{fig:compareB}
\end{figure}

We evaluate physical landslide event detection with LITMUS-ASSED and compare to the LITMUS-original application from ~\cite{r_esa,litmus_a} in Figure~\ref{fig:compareA} and ~\ref{fig:compareB}. 

Figure~\ref{fig:compareB} shows normalized additional events detected by LITMUS-ASSED and LITMUS-original. In each data window, LITMUS-ASSED detects significantly more events than the static LITMUS-original application (\texttt{N\_RES}). Additionally, under our \texttt{RES-HDI} approach, increasing percentages of true physical-event detections are performed by the LITMUS-ASSED application, compared to LITMUS-original. By December 2018, LITMUS-ASSED detects 370 additional events, while LITMUS-original only detects 149 events (these 149 events are also detected in LITMUS-ASSED), bringing the total in LITMUS-ASSED to 519 events. This is an increase of 348.3\% in LITMUS-ASSED compared to LITMUS-original. LITMUS-ASSED is also able to correct wrong detections made by LITMUS-original, decreasing false positives and increasing precision. Details for each window are provided in Table~\ref{tab:detections} (LITMUS-ASSED's accuracy has already been compared to LITMUS-original in Figures~\ref{fig:results} with labeled data; this table only covers true physical events).

\begin{table}[h]
	\caption{\texttt{False pos.} are false-positive landslide detections in LITMUS-original that LITMUS-ASSED correctly identifies as irrelevant. \texttt{False neg.} are events LITMUS-original does not detect, but are correctly detected as landslides in LITMUS-ASSED. \texttt{$\sum$ LITMUS-ASSED} is the total events detected by LITMUS-ASSED (sum of \texttt{Both apps.} and \texttt{False neg.})}
	\label{tab:detections}
	\begin{tabular}{l r r r r r r }
		& Jul & Aug & Sept & Oct & Nov & Dec \\ 
		\hline 
		Both apps. & 480 & 644 & 365 & 501 & 508 & 149 \\ 
		False pos. & 44 & 75 & 20 & 40 & 27 & 5 \\ 
		False neg. & 398 & 681 & 513 & 646 & 772 & 370 \\ 
		\hline 
		$\sum$ LITMUS-ASSED & 878 & 1325 & 878 & 1147 & 1280 & 519 \\ 
		Pct Increase & 183\% & 206\% & 241\% & 229\% & 252\% & 348\% \\ 
		\hline 
	\end{tabular} 
\end{table}

%
%

\section{Conclusions}
\label{sec:conc}
We proposed ASSED, an adaptive social sensor event detection framework designed for physical event detection on social media streams. ASSED can handle heterogenous data sources such as numeric sensors, reputable sources, and social media data. ASSED’s dataflow is designed for fast prototyping and deployment, with fault-tolerant decoupled process-to-process communication. ASSED’s dataflow allows applications to handle variable data ingest frequencies, and bursty streams. Through ASSED, we presented the following:

\begin{enumerate}
	\item ASSED includes an ML-based  event processing engine that continuously adapts to changes in \textit{live data}. We examined ASSED's filter generation, update, and prediction methods. ASSED's ML-based event detector is able to identify physical events from social media streams and augment low-coverage high-confidence sources with high-coverage global social sensors.
	\item ASSED's Heterogenous Data Integration (HDI) process integrates high-confidence sources with social streams. HDI forms a core component of ASSED's adaptivity. We demonstrate HDI's efficacy through improvements in physical event detection in ASSED compared to static approaches.
	\item We believe the ASSED framework can be useful for a variety of social-sensor based physical event detection. We demonstrated a landslide detection application LITMUS-ASSED built on ASSED that improves upon static approaches such as LITMUS-original. LITMUS-ASSED adapts to changing event characteristics in social sources and detects almost 350\% more landslide events than LITMUS-original. Moreover, LITMUS-ASSED maintains high-fidelity event detection accuracy, with \textit{f-score} of 0.988 by December 2018, compared to \textit{f-score} of 0.762 for static approaches (LITMUS-original).		
\end{enumerate}

ASSED has a few avenues for future work. 
Currently, ASSED will store the entire training data for a filter in \textit{F\_Store} for future retrieval. Using \textit{k}-NN on large datasets is computationally expensive, increasing delays in event detection in the Hybrid filter selection approach in the Event Detection process. We can consider approximate nearest neighbor searches with locality sensitive hashing ~\cite{lsh_andoni} as ways to improve retrieval speed.


%

%
\begin{acks}
This research has been partially funded by National Science Foundation by CISE’s SAVI/RCN (1402266, 1550379), CNS (1421561), CRISP (1541074), SaTC (1564097) programs, an REU supplement (1545173), and gifts, grants, or contracts from Fujitsu, HP, Intel, and Georgia Tech Foundation through the John P. Imlay, Jr. Chair endowment. Any opinions, findings, and conclusions or recommendations expressed in this material are those of the author(s) and do not necessarily reflect the views of the National Science Foundation or other funding agencies and companies mentioned above.
\end{acks}

%
\bibliographystyle{ACM-Reference-Format}
\bibliography{main}

\end{document}